# The effect of compressive strain on the Raman modes

# of the dry and hydrated BaCe$_{0.8}$Y$_{0.2}$O$_3$ proton conductor

Qianli Chen[1,2,*], Tzu-Wen Huang[1], Maria Baldini[3,4], Anwar Hushur[5],

Vladimir Pomjakushin[6], Simon Clark[7], Wendy L. Mao[3,4], Murli H. Manghnani[5],

Artur Braun[1,8,*], Thomas Graule[1,9]

[1]*Laboratory for High Performance Ceramics*
*Empa. Swiss Federal Laboratories for Materials Science and Technology*
*CH-8600 Dübendorf, Switzerland*

[2] *Department of Physics, ETH Zürich, Swiss Federal Institute of Technology*
*CH-8057 Zürich, Switzerland*

[3]*Geological and Environmental Sciences, Stanford University,*
*Stanford, California 94305, USA*

[4] *Stanford Institute for Materials and Energy Science*
*SLAC National Accelerator Laboratory, Menlo Park, California 94025, USA*

[5]*Hawaii Institute of Geophysics, University of Hawaii at Manoa*
*Honolulu, Hawaii 96822, USA*

[6] *Laboratory for Neutron Scattering, Paul Scherrer Institut*
*CH-5232 Villigen PSI, Switzerland*

[7] *Advanced Light Source, LBNL, 1 Cyclotron Road, Berkeley, CA 94720, USA*

[8]*Hawaii Natural Energy Institute, University of Hawaii at Manoa*
*Honolulu HI 96822, USA*

[9] *Institute of Ceramic, Glass- and Construction Material,*
*TU Bergakademie Freiberg, D-09596 Freiberg, Germany*

*Corresponding Author: Phone +41 44 823 4129; +41 44 823 4850, Fax: +41 44 823 4150,

E-mail: qianlichen1@gmail.com; artur.braun@alumni.ethz.ch

ABSTRACT

The $BaCe_{0.8}Y_{0.2}O_{3-\delta}$ proton conductor under hydration and under compressive strain has been analyzed with high pressure Raman spectroscopy and high pressure x-ray diffraction. The pressure dependent variation of the $A_g$ and $B_{2g}$ bending modes from the O-Ce-O unit is suppressed when the proton conductor is hydrated, affecting directly the proton transfer by locally changing the electron density of the oxygen ions. Compressive strain causes a hardening of the Ce-O stretching bond, with the pressure coefficient $\Delta v/\Delta p = 4.32\pm0.05$ $cm^{-1}$/GPa being the same for the dry and hydrated sample. As a result of this hardening of the lattice vibrations, the activation barrier for proton conductivity is raised, in line with recent findings using high pressure and high temperature impedance spectroscopy. Hydration also offsets slightly the Ce-O $B_{1g}$ and $B_{3g}$ stretching modes by around 2 $cm^{-1}$ towards higher wave numbers, revealing an increase of the bond strength of Ce-O. The (20-2) Bragg reflections do not change during pressurizing and thus reveal that the oxygen occupying the O2 site displaces only along the *b*-axis. The increasing Raman frequency of the $B_{1g}$ and $B_{3g}$ modes thus implies that the phonons become hardened and increase the vibration energy in the *a-c* crystal plane upon compressive strain, whereas phonons are relaxed in the *b*-axis, and thus reveal softening of the $A_g$ and $B_{2g}$ modes. Lattice toughening in the *a-c* crystal plane raises therefore a higher activation barrier for proton transfer and thus anisotropic conductivity. Particularly for the development of epitaxial strained proton conducting thin film devices with lower activation energy, such anisotropy has to be taken quantitatively into account. The experimental findings of the interaction of protons with the ceramic host lattice under external strain may provide a general guideline for yet to develop epitaxial strained proton conducting thin film systems with high proton mobility and low activation energy.

## 1. Introduction

Ceramic proton conductors have attracted attention for their potential applications as gas sensors and as electrolytes in solid oxide fuel cells and electrolyzers.[1-3] $A^{2+}B^{4+}O^{2-}_3$ –type perovskite oxide ceramics, such as barium cerates and zirconates ($BaCeO_3$, $BaZrO_3$) substituted by trivalent $B^{3+}$ rare earth ions, have been thoroughly investigated because of their high proton conductivity.[4, 5] The substitution is coupled with oxygen vacancy formation in order to maintain charge balance. Upon exposure to moisture the proton conductors absorb the water molecules, the hydroxyl groups of which fill the oxygen vacancies, whereas the remaining protons are situated in the crystal lattice. At temperatures typically ranging from 623 K to 823 K the proton diffusion sets on. At about this temperature range we found that the thermal expansion of hydrated, Y-substituted barium zirconate has a discontinuity, which is absent in the dry material.[6] The mechanism for proton diffusion has been suggested as the hopping of protons between neighboring oxygen ions in the presence of hydroxyl groups.[7-9] Raman scattering has been utilized to probe the influence of protons on the elastic properties (though with lower sensitivity than infrared spectroscopy),[10] and more importantly, the bonds of the host lattice, thus improving understanding of the interactions of protons with the crystal lattice.[11-14] Recent Raman spectroscopy studies reported small wavenumber shifts of 1 $cm^{-1}$ - 5 $cm^{-1}$ towards higher wave numbers after water exposure in Y or Yb doped $BaZrO_3$ and $SrZrO_3$,[13, 15] revealing a change of the covalence of bonds and the polarizability of the Zr atoms by the protons. We confirm such small shift in the Raman spectra also for $BaCe_{0.8}Y_{0.2}O_{3-\delta}$.

Pressure dependent studies can provide additional insight into the interaction of protons with the host lattice, since the application of external pressure – tensile or compressive – represents an additional thermodynamic variable which can be used to tune the spatial proximity between proton and host and thus their interactions.[16, 17] Earlier implications that the proton conductivity activation energy $E_a$ can be lowered by expanding the lattice volume - using chemical pressure by either using cation substitution or different synthesis methods[18] - were directly confirmed by our high pressure and temperature dependent impedance studies.[19, 20] High pressure Raman spectroscopy and X-ray and neutron diffraction of hydrated and non-hydrated proton conductors can shed light on the influence of structural or vibrational properties on the proton transport. For further insight into the causal relationship between the lattice volume changes and the proton transport properties, we investigated the high-pressure crystal structure and phonon vibrations of $BaCe_{0.8}Y_{0.2}O_{3-\delta}$ not only in the dry but also also, to the best of the authors knowledge for the first time, in the hydrated states.

## 2. Experimental

$BaCe_{0.8}Y_{0.2}O_{3-\delta}$ powder was synthesized by conventional solid-state reaction as detailed in ref. 19. The powders and pellets annealed for 15 h at 973 K in dry $O_2$ and cooled afterwards are termed "dry"; powder and pellets heated in $O_2$ saturated with $H_2O$ to 723 K for 8 h are termed "hydrated". Pressure dependent X-ray diffraction was carried out at the beamline 12.2.2 at the Advanced Light Source (ALS), Lawrence Berkeley National Laboratory, using an X-ray wavelength of 0.496 Å.[21] Polycrystalline samples were loaded into a sample chamber drilled in CuBe gasket, with a ruby chip for pressure calibration. The gasket was compressed in a diamond anvil cell (DAC). Diffraction patterns were recorded with a 2D detector. The 2D images were collected using a Mar345 image plate and processed using Fit2d software. The data was calibrated using a $LaB_6$ standard.[22] Neutron powder diffractograms (NPD) for the deuterated sample were recorded with wavelength $\lambda = 1.9$ Å at the HRPT diffractometer beamline, SINQ Neutron Spallation Source.[23] Structure parameters were refined using Rietveld analysis with GSAS software package. [24] Optical Raman spectra for the dry and hydrated sample were recorded at ambient conditions with a Dilor XY spectrometer equipped with a liquid nitrogen-cooled charge-coupled-device (CCD) detector using the 514.5 nm green line of a Spectra Physics Ar ion laser. An objective with magnification 50x was used to focus the incident laser light on the sample, the spot size is about 3 μm diameter, and to collect the light. All spectra were recorded in the backscattering geometry with no polarization used for the collected signal. The laser power is 30 mW on the sample. The spectrometer was calibrated using single-crystal silicon as a reference. High pressure Raman spectra were recorded in a DAC with a Renishaw RM1000 Raman microscope using a 514-nm laser excitation line. [25] The authors are not aware of any previous high pressure Raman spectroscopy and x-ray diffraction study of hydrated ceramic proton conductors.

## 3. Results and Discussions

### 3.1 Effect of hydration in bond strength

Raman spectra of the dry and hydrated $BaCe_{0.8}Y_{0.2}O_{3-\delta}$ (Figure 1) collected under ambient conditions show two prominent resonances at 465 cm$^{-1}$ and at 3560 cm$^{-1}$. The resonance at 3560 cm$^{-1}$ is only visible for the hydrated sample and is attributed to the hydroxyl (OH-) groups which are formed in the perovskite upon hydration.[10] The spectrum of the dry $BaCe_{0.8}Y_{0.2}O_{3-\delta}$ has an extra pronounced shoulder at 465 cm$^{-1}$ which we attribute to a vibration related with oxygen vacancies.[15, 26] Both peaks are thus considered conjugated spectral signatures for the filling of oxygen vacancies with hydroxyl groups.

The upper inset in Figure 1 highlights the spectral range from 300 – 400 cm$^{-1}$ which we have deconvoluted into several resonant bands. Bands in the range from around 310 to 350 cm$^{-1}$ have recently been spectrally assigned to $BaCeO_3$ [16, 27]. In parallel, we consider this assignment here valid for $BaCe_{0.8}Y_{0.2}O_{3-\delta}$ as follows: band $v_1$ at about 310 cm$^{-1}$ (labeled C in ref. [16]) is a resonance with $B_{2g}$ symmetry and originates from Ba-O stretching (≥50%) and O-Ce-O bending mode.[27] Band $v_2$ at around 330 cm$^{-1}$ (labeled B in ref. [16]) contains the resonances with $A_g$ and $B_{2g}$ symmetry, where $A_g$ dominates and involves also Ba-O stretching ($\geq$30%) and O-Ce-O bending, which can be concerned similar as $v_i$.[27] Band $v_3$ at around 355 cm$^{-1}$ (labeled A in ref. [16]) is a superposition of resonances with $B_{2g}$, $A_g$, $B_{3g}$ and $B_{1g}$ symmetry which can be resolved only at temperatures as low as 77 K. The two latter resonances are the dominant ones and arise from pure Ce-O stretching.[27]

The Raman spectra of dry and hydrated $BaCe_{0.8}Y_{0.2}O_{3-\delta}$ are compared for ambient pressure and pressures up to 8.4 GPa in the range of 130 – 500 cm$^{-1}$ in Figure 2. Overall, it appears that the spectra of hydrated $BaCe_{0.8}Y_{0.2}O_{3-\delta}$ have somewhat sharper features than the spectra from dry $BaCe_{0.8}Y_{0.2}O_{3-\delta}$. For further quantitative analyses, particularly for the pressure dependent studies, we have deconvoluted the spectral range from 200 to 500 cm$^{-1}$ into Gaussians so as to be able to determine peak positions for the resonances $v_1$, $v_2$ and $v_3$ accurately by least square fitting.

Based on literature data of $BaCeO_3$, [16, 27] the Raman vibration modes in Figure 2 are identified as the following: For p < 3 GPa, a Ce-O stretching peak overlaps with O-Ce-O bending modes at 351 cm$^{-1}$ due to strong parasitic scattering from the diamond anvil cell. For the hydrated sample at p=3.5 GPa, $v_3$ shifts to higher frequency and becomes visible at 387 cm$^{-1}$ (Figure. 2b).

The splitting of the Raman modes upon compression suggests lowering of the cell symmetry and increase of the multiplicity of the lattice parameters. Both the dry and hydrated samples show shifts to higher wavenumbers upon pressurising. For the shift from the Ce-O stretching modes $v_3$, the hydrated sample shows slightly larger frequency shifts for all the pressures measured, and the splitting of peaks was more pronounced for the hydrated ones than for the dry ones. In contrast, for the O-Ce-O bending modes $v_1$ and $v_2$, the evolution of the wavenumbers were small and almost identical for both the dry and hydrated cases.

Figure 3 shows the evolution of the wavenumbers for the Raman-active bands with increasing pressure. Visual inspection of the general trends in Figure 3 shows that $v_3$ (Ce-O stretching mode) increases linearly with applied pressure from 373 cm$^{-1}$ to over 400 cm$^{-1}$, whereas $v_1$ and $v_2$ remain more or less constant at 312 cm$^{-1}$ and 350 cm$^{-1}$, respectively. Linear least square fits for the data from the dry and hydrated samples show the same slope with an average pressure coefficient $\Delta v_3/\Delta p = 4.32\pm0.05$ cm$^{-1}$/GPa, but an intercept by 2.1 cm$^{-1}$ higher for the hydrated sample in peak $v_3$. Since the slope is the same, we assume that the increase of the bond strength for the dry and hydrated $BaCe_{0.8}Y_{0.2}O_{3-\delta}$ occurs homogeneously. A weak shift of about 4 cm$^{-1}$ wavenumbers upon hydration has been found before at ambient conditions in ref. [14]; parallel to this the opposite shift after thermal dehydration was observed as well, which can be attributed to a characteristic change when adding water to the proton conductor lattice.

The pressure dependence of these three Raman active bands $v_1$, $v_2$ and $v_3$ is an indication of the $CeO_6$ octahedron deformation, see also ref. [16]. The force constant of the Ce-O stretching is much more sensitive to the bond length variation than the Ba-O stretching, and also to the shearing modes, thus band $v_3$ is more sensitive towards pressure than $v_1$ and $v_2$.[27] The relative small variations of $v_1$ and $v_2$ from the O-Ce-O bending correspond to the shearing of the octahedra.

When the bond situation changes in response to *hydration*, this will reflect in the vibrational structure as measured with Raman spectroscopy. Because the shift of the Raman resonances is in the direction towards larger wavenumbers, we conclude that hydration increases the Ce-O stretching bond strength. Consequently, hydration of $BaCe_{0.8}Y_{0.2}O_{3-\delta}$ comprises also an increase of the hardness. It remains open whether the increase of hardness is a result of oxygen vacancy filling or a result of hydrogen interstitials, but since hydrogen bonds are known to be strong, latter scenario seems the more likely one. The latter effect is also the reason for hy-dride bond formation in metals, for examples, and manifests in mechanical brittleness and thus shows that it is the protons during hydration which cause the increase of the Ce-O stretching bond.

We also notice that in the hydrated sample in Figure 3, $v_1$ and $v_2$ show smaller pressure dependent variations in Raman frequency, meaning that the O-Ce-O bending force constants are more stable in response to pressure after hydration. Since hydration goes along with the formation of hydroxyl groups into the oxygen vacancies, the role of hydrogen can be considered as to prevent the hardening of O-Ce-O bending force constants. Since the bending modes of O-Ce-O lead to the variations of electron density of neighboring oxygen ions,[28] "flexible" bending force constants allow facilitate proton transfer.

The above behavior in the stretching and bending of $CeO_6$ is fully in line with our recent finding from high temperature dependent neutron diffraction that in the proton conducting perovskites, hydration causes a decrease in the lattice volume,[6] in response to hydrogen bond formation.

## 3.2 Effect of pressure in bond strength, lattice structure and proton transfer

We have found recently that the proton transport activation energy $E_a$ increases with the applied compressive strain.[20] The vibration of ions in the host lattice of proton conductor can be pictured as an oscillation. The activation energy $E_a$ for proton conductivity is known to scale with the relative displacement $x$ of the oscillators and thus with the Raman resonance vibration frequency $v$:[20]

$$E_a \propto (1/x^2) \propto v^2 \qquad (1)$$

With respect to the aforementioned applications of proton conductors such as for solid oxide fuel cell or electrolyzer electrolytes, a lower proton transport activation energy is desirable. This could be technically achieved by applying tensile strain to the proton conductor as an epitaxial strained thin film,[28] such as suggested in Figures 4 and 5 in ref. [19]. Because transverse contraction would cause compressive strain in the perpendicular direction, the activation energy would increase in the direction of lateral proton transport, however, so that a specific tailoring of film growth conditions is necessary to warrant facilitated proton transport in a particular direction.

Taking into account the results from the Raman modes analysis, the influence of pressure in $E_a$ can be further explained as a consequence of the shift in modes $v_3$, which represents Ce-O stretching. In other words, compressive strain causes stronger Ce-O bonds and thus lattice hardening and less lattice vibration, which may keep the protons in the $CeO_6$ octahedral cage, and raises therefore the energy barrier for the proton transfer.

The changes in the Raman modes upon pressurizing the $BaCe_{0.8}Y_{0.2}O_{3-\delta}$ originate from the changes in the bond lengths, i.e. the crystal lattice parameters or interatomic distances. The high pressure X-ray diffractograms for hydrated $BaCe_{0.8}Y_{0.2}O_{3-\delta}$ (not shown here) reveal that the pressure induces a decrease in the lattice parameters and a decrease in the unit cell volume in the pressure range from ambient to 9 GPa, as displayed in Figure 4a. This structural change is paralleled by an increase in the Ce-O stretching force constants for the dry and hydrated $BaCe_{0.8}Y_{0.2}O_{3-\delta}$, see Figure 3. No phase transition has been reported for the range of pressure applied, [16] therefore we assume that the lattice parameters decrease homogeneously upon pressurizing.

Structure refinement of the high pressure X-ray powder diffraction data allows us to detail the fractional position of the atoms in the $BaCe_{0.8}Y_{0.2}O_{3-\delta}$. In Figure 4b, the $\beta$ angle, which is defined as the angle including the $a$-axis and the $c$-axis, increases proportional with pressure with a slope of $0.0798°$ per GPa. Correspondingly, the volume of the orthorhombic phase was calculated as a function of the pressure in Figure 4a. For reference we have also included the structure data point (from neutron diffraction) for the hydrated sample after the pressure had been released (relaxed state). The neutron diffractogram in Figure 5a shows how well we have refined the structure of the $BaCe_{0.8}Y_{0.2}O_{3-\delta}$ under ambient conditions.

Figure 5b shows a portion of the high pressure X-ray diffractograms (open circle) with structure refinement curves (solid line) as pressure dependence in the (20-2), (040), (123) and the (321) Bragg reflections. The peak intensity decreases at (040), (123) and (321) and can be rationalized and simulated by local movement of oxygen ions occupying the crystallographic O2 site. Refer to Figure 6, where we have plotted the positions of the Ba, Ce and O atoms for better visualization. However, the intensity of the (20-2) reflections is invariant versus pressure change. The reason for this invariance is that the O2 site with miller indices (0, b, 0) is a Bragg forbidden peak intensity contribution if O2 displaces only along the *b*-axis direction when the pressure is increasing. Furthermore, in this assumption, all atoms other than O2 should keep the fractional coordinates fixed during the lattice changes as a function of pressure. Since the constant intensity of the (20-2) reflections upon increasing pressure imply that all other atoms - with the exception of O2 - keep the fractional coordinates fixed, we may ignore the intensity of the O3 contribution in the Raman spectra, the O3 site of which is the other oxygen site linking two Ce centered octahedra in the *b*-axis. Let us inspect the pressure dependent Raman modes in Figure 3: the $B_{1g}$ mode, O2 atom displacement at (0.349 0 0.357), increases linearly as function of pressure. But the $A_g$ mode with O2 atom displacement at (0.0.146 -0.212 0.2), and the $B_{2g}$ mode with O2 atom displacement at (-0.355 0.064 0.346) remain virtually constant upon pressurizing. Since the O2 atom is located at the nexus between the Ce atoms and sharing atoms at two Ce centered octahedral at the *a-c* plane, as Figure 5 shows, the O2 position directly affects the bond strength and the phonon behavior between the Ce-O2-Ce nexus. When we relate this with the virtually absent intensity change in the $A_g$, $B_{2g}$ modes, the increasing phonon energy of the $B_{1g}$ mode should imply that the phonon becomes hardened and increases the vibration energy in the *a-c* plane as pressure in-creasing; whereas the phonon energies are relaxed in the *b*-axis as the O2 position and the angle enclosed by Ce-O2-Ce (assigned as $\gamma$) are mutually adjusting.

Figure 7 shows the variation of $\gamma$ and the bond length of Ce-O2 as a function of pressure as obtained by the Rietveld refinement of the XRD data. The decrease of $\gamma$ along with the increasing bond distance of Ce-O2 upon increasing the pressure give an acceptable rational for the softening of the $A_g$, $B_{2g}$ phonon modes. This finding indicates that the lattice hardening at the *a-c* crystal plane may raise a higher barrier for the proton transfer in the lattice, i.e. the proton transport may be anisotropic in the $BaCe_{0.8}Y_{0.2}O_{3-\delta}$ lattice. Anisotropic conductivity may have particularly implications for the aforementioned epitaxial strained proton conducting films, because the film growth conditions for the proton conductor would depend on the choice of the substrate material and orientation. Depending on whether fast proton transport should be in the lateral or perpendicular film direction, the *a-c* crystal plane should grow accordingly as well. Specifically, for fast proton transport perpendicular to the film place, the *a-c* plance with higher activation energy should be in perpendicular direction to the film plane, for example.

When we turn again to Figure 3, we notice a small discontinuity at around 3 – 4 GPa in the pressure dependence of the Raman modes, which possibly implies a small distortion in the structure at this pressure. This suggestion however could not be confirmed within the detection limits of XRD.

Raman spectroscopy provides only the partial density of states in the Raman active range. For a deeper understanding of the proton dynamics in $BaCe_{0.8}Y_{0.2}O_{3-\delta}$, the interaction between two adjacent $CeO_6$ octahedra should be studied. To describe the phonon behavior for each octahedron, not only the phonon modes of each atom need to be understood, but also the energy distribution and vectors of each modes, which requires measurements of the phonon distribution, such as with inelastic x-ray scattering or inelastic neutron scattering. Extending those epxeriments plus Raman spectroscopy to high temperature and combining these with structural studies will likely benefit the understanding of correlation between the near-neighborhood structure and the proton transport.

## 4. Conclusion

The effect of compressive strain and hydration on the $BaCe_{0.8}Y_{0.2}O_{3-\delta}$ proton conductor at ambient temperature has been investigated with high pressure optical Raman spectroscopy in combination with neutron diffraction and high pressure x-ray diffraction. Hydration of $BaCe_{0.8}Y_{0.2}O_{3-\delta}$ suppresses the variation and thus stabilizes the O-Ce-O $A_g$ and $B_{2g}$ bond bending modes upon pressing, which affect the proton transfer by changing the electron density locally. Hydration also enhances slightly the Ce-O $B_{1g}$ and $B_{3g}$ stretching modes towards higher wave numbers, revealing an increase of the bond strength of Ce-O.

Application of compressive strain causes a hardening of the Ce-O stretching bond, with the pressure coefficient $\Delta v_3/\Delta p = 4.32\pm0.05$ cm$^{-1}$/GPa being the same for the dry and hydrated sample. As a result of this hardening of the lattice vibrations by compressive strain, the activation barrier for proton conductivity is also raised, and this is in line with our recent experimental findings using high pressure and high temperature impedance spectroscopy.

Meanwhile, the invariant intensity of the X-ray (20-2) Bragg reflections versus pressure change reveals that the oxygen occupying the O2 site displaces only along the *b*-axis. Therefore, the increasing Raman frequency of the $B_{1g}$ and $B_{3g}$ modes implies that the lattice phonons become hardened and increase the vibration energy in the *a-c* crystal plane upon pressurizing, whereas they are relaxed in the *b*-axis, and thus revealing softening of the $A_g$ and $B_{2g}$ modes. The lattice toughening in the *a-c* crystal plane may thus raise a higher activation barrier for the proton transfer in the lattice; in so far, the proton transport may be anisotropic in the $BaCe_{0.8}Y_{0.2}O_{3-\delta}$ lattice. Particularly for the development of epitaxial strained proton conducting thin film devices, such anisotropy has to be taken quantitatively into account.

**Acknowledgement**

Financial support for Q.C. by Swiss National Science Foundation Grant # SNF 200021-124812, for A.B. by SNF # IZK0Z2-133944, for T.-W. H. by European Union SOFC-Life project # 256885, and for A.H. and M.M. by the U.S. National Science Foundation grant # 0538884 is gratefully acknowledged. Technical assistance at high pressure XRD by Edvinas Navickas (Kaunas University of Technology) is gratefully acknowledged. The ALS is supported by the Director, Office of Science, Office of Basic Energy Sciences, of the U.S. Department of Energy under Contract No. DE-AC02-05CH11231. This work is based on experiments performed at the Swiss Spallation Neutron Source SINQ, Paul Scherrer Institut, Villigen, Switzerland.

**References**

(1)     Kreuer, K. D. *Annu. Rev. Mater. Res.* **2003**, *33*, 333-359.

(2)     Norby, T. *Proton conductivity in perovskite oxides*. In *Perovskite oxide for solid oxide fuel cells*; Ishihara, T., Eds.; (Springer, 2009), pp. 217-241.

(3)     Iwahara, H. *Proton Conductors: Solids, membranes and gels - materials and devices*. University Press: Cambrigde, 1992.

(4)     Iwahara, H. *Solid State Ionics* **1995,** *77*, 289-298.

(5)     Coors, W. G.; Readey, D. W. *J. Am. Ceram. Soc.* **2002,** *85*, 2637-2640.

(6)     Braun, A.; Ovalle, A.; Pomjakushin, V.; Cervellino, A.; Erat, S.; Stolte, W. C.; Graule, T. *Appl. Phys. Lett.* **2009,** *95*, 224103.

(7)     Hempelmann, R.; Karmonik, C.; Matzke, T.; Cappadonia, M.; Stimming, U.; Springer, T.; Adams, M. A. *Solid State Ionics* **1995,** *77*, 152-156.

(8)     Norby, T.; Widerøe, M.; Glöckner, R.; Yngve, L. *Dalton Trans.* **2004,** *19*, 3012-3018.

(9)     Islam, M. S.; Davies, R. A.; Gale, J. D. *Chemical Communications* **2001**, 661-662.

(10)    Glerup, M.; Poulsen, F. W.; Berg, R. W. *Solid State Ionics* **2002,** *148*, 83-92.

(11)    Scherban, T.; Villeneuve, R.; Abello, L.; Lucazeau, G. *Solid State Ionics* **1993,** *61*, 93-98.

(12)    Karlsson, M.; Matic, A.; Knee, C. S.; Ahmed, I.; Eriksson, S. G.; Börjesson, L. *Chem. Mater.* **2008,** *20*, 3480-3486.

(13)    Karlsson, M.; Matic, A.; Berastegui, P.; Börjesson, L. *Solid State Ionics* **2005,** *176*, 2971-2974.

(14)    Slodczyk, A.; Colomban, P.; Lamago, D.; Limage, M.-H.; Romain, F.; Willemin, S.; Sala, B. *Ionics* **2008,** *14*, 215-222.

(15)    Slodczyk, A.; Colomban, P.; Willemin, S.; Lacroix, O.; Sala, B. *J. Raman Spectrosc.* **2009,** *40*, 513-521.

(16)    Loridant, S.; Lucazeau, G. *J. Raman Spectrosc.* **1999,** *30*, 485-492.

(17)    Zhang, J.; Zhao, Y.; Xu, H.; Li, B.; Weidner, D. J.; Navrotsky, A. *Elastic properties of yttrium-doped BaCeO3 perovskite*. AIP: 2007.

(18)    Duval, S. Y-substituted barium zirconate, a proton conducting electrolyte for applications at intermediate temperatures. Techn. Univ. München, 2008.

(19)    Chen, Q.; Braun, A.; Ovalle, A.; Savaniu, C.-D.; Graule, T.; Bagdassarov, N. *Appl. Phys. Lett.* **2010,** *97*, 041902-041903.

(20)    Chen, Q.; Braun, A.; Yoon, S.; Bagdassarov, N.; Graule, T. *Journal of the European Ceramic Society* **2011,** *31*, 2657-2661.

(21)   Kunz, M.; MacDowell, A. A.; Caldwell, W. A.; Cambie, D.; Celestre, R. S.;
       Domning, E. E.; Duarte, R. M.; Gleason, A. E.; Glossinger, J. M.; Kelez, N.; Plate, D.
       W.; Yu, T.; Zaug, J. M.; Padmore, H. A.; Jeanloz, R.; Alivisatos, A. P.; Clark, S. M.
       *J. Synchrotron Radiat.* **2005,** *12*, 650-658.

(22)   Yan, J.; Knight, J.; Kunz, M.; Vennila Raju, S.; Chen, B.; Gleason, A. E.; Godwal, B.
       K.; Geballe, Z.; Jeanloz, R.; Clark, S. M. *J. Phys. Chem. Solids* **2010,** *71*, 1179-1182.

(23)   Fischer, P.; Frey, G.; Koch, M.; Könnecke, M.; Pomjakushin, V.; Schefer, J.; Thut,
       R.; Schlumpf, N.; Bürge, R.; Greuter, U.; Bondt, S.; Berruyer, E. *Phys. B.* **2000,** *276-
       278*, 146-147.

(24)   Larson, A. C.; Dreele, R. B. V. *General Structure Analysis System (GSAS).*  Los
       Alamos National Laboratory Report LAUR 86-748, 2000.

(25)   Lin, Y.; Mao, W. L.; Mao, H.-k. *Proc. Natl. Ac. Sci.* **2009,** *106*, 8113-8116.

(26)   Mineshige, A.; Okada, S.; Kobune, M.; Yazawa, T. *Solid State Ionics* **2007,** *178*, 713-
       715.

(27)   Genet, F.; Loridant, S.; Lucazeau, G. *J. Raman Spectrosc.* **1997,** *28*, 255-276.

(28)   Santiso, J.; Burriel, M. *Journal of Solid State Electrochemistry* **2011,** *15*, 985-1006.

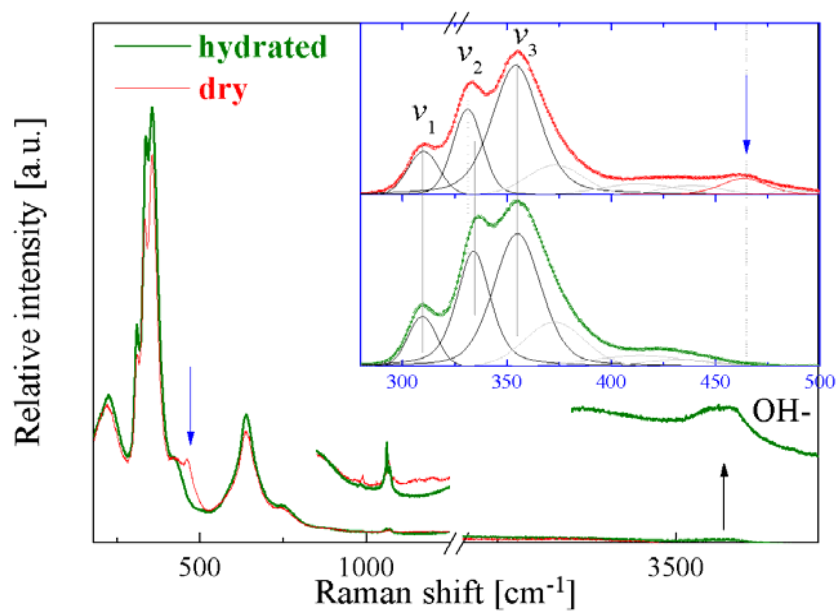

Figure 1. Raman spectra of dry and hydrated BaCe$_{0.8}$Y$_{0.2}$O$_{3-\delta}$ at ambient condition.

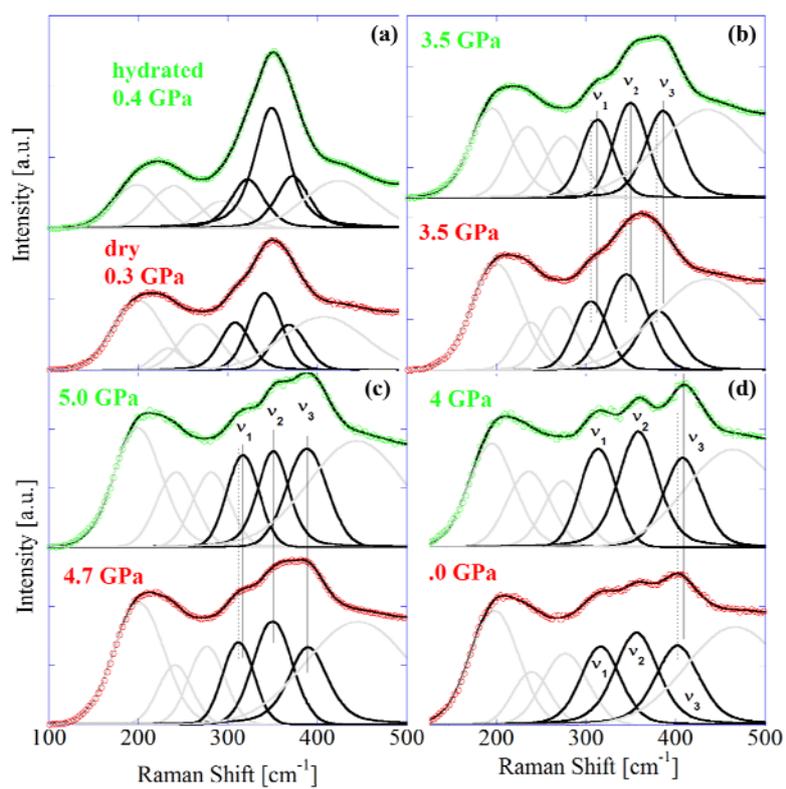

Figure 2. Raman spectra of dry and hydrated BaCe$_{0.8}$Y$_{0.2}$O$_{3-\delta}$ upon pressurizing. (a) 0.3 − 0.4 GPa; (b) 3.5 GPa; (c) 4.7 - 5.0 GPa; (d) 8.0 − 8.4 GPa.

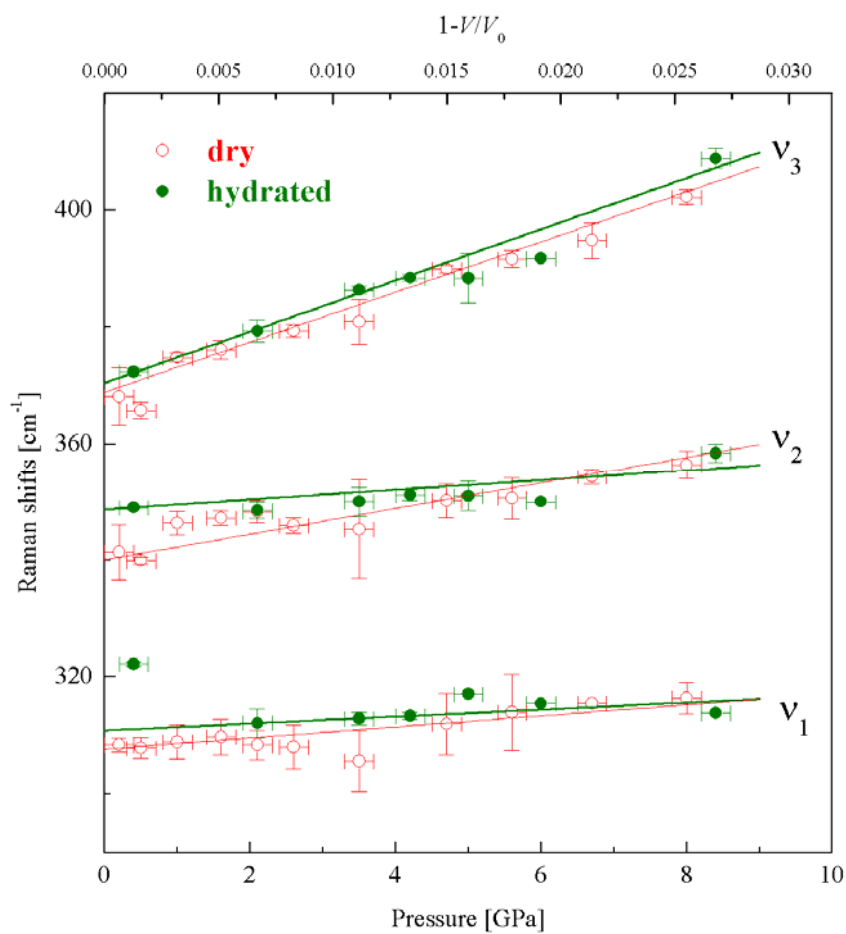

Figure 3. Raman frequency shifts as a function of pressure.

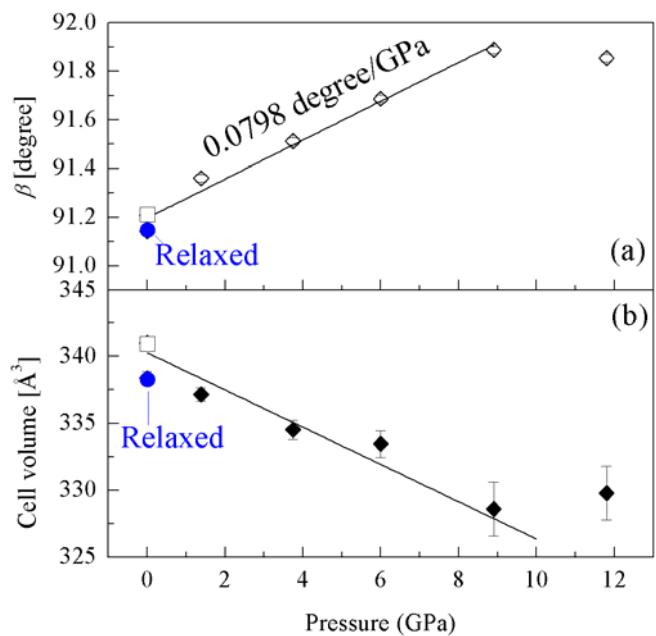

Figure 4. Refined lattice parameters for hydrated BaCe$_{0.8}$Y$_{0.2}$O$_{3-\delta}$ as a function of pressure:

the unit cell volume (a) and the β angle (b).

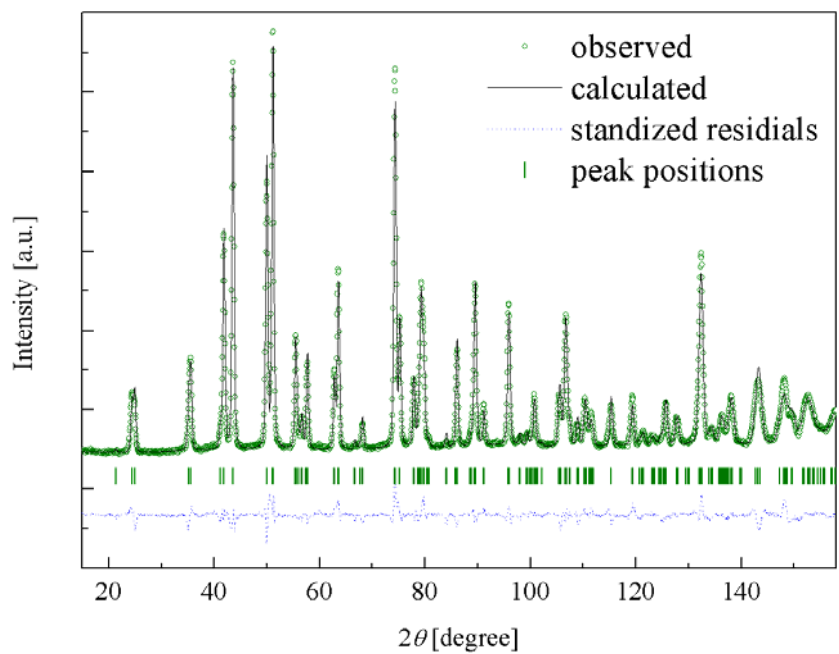

Figure 5. Neutron diffraction pattern for deuterated BaCe$_{0.8}$Y$_{0.2}$O$_{3-\delta}$ at ambient condition (a) and fragment of X-ray diffraction patterns and refinements for hydrated BaCe$_{0.8}$Y$_{0.2}$O$_{3-\delta}$ under various pressures (b).

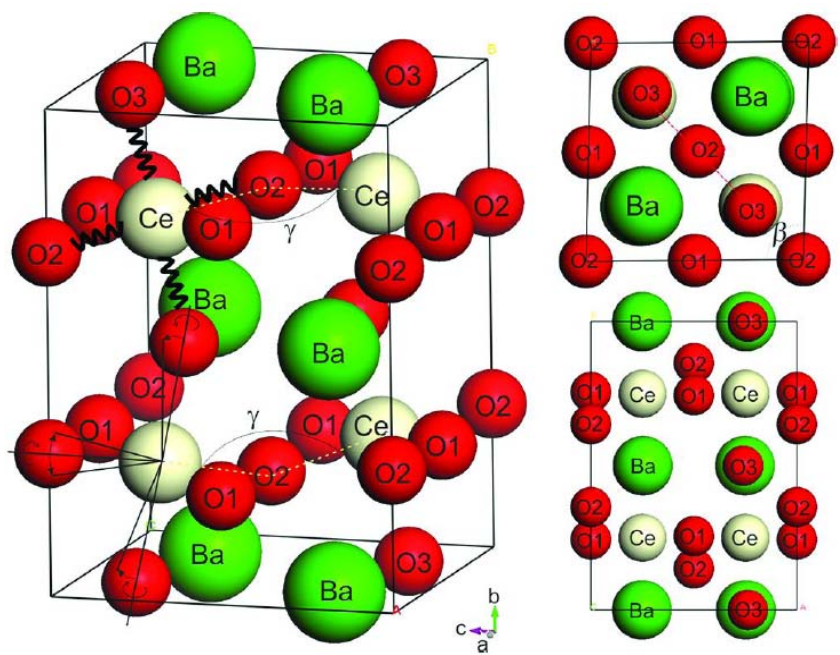

Figure 6. Sketch of crystal structure for hydrated BaCe$_{0.8}$Y$_{0.2}$O$_{3-\delta}$ at 1.39 GPa.

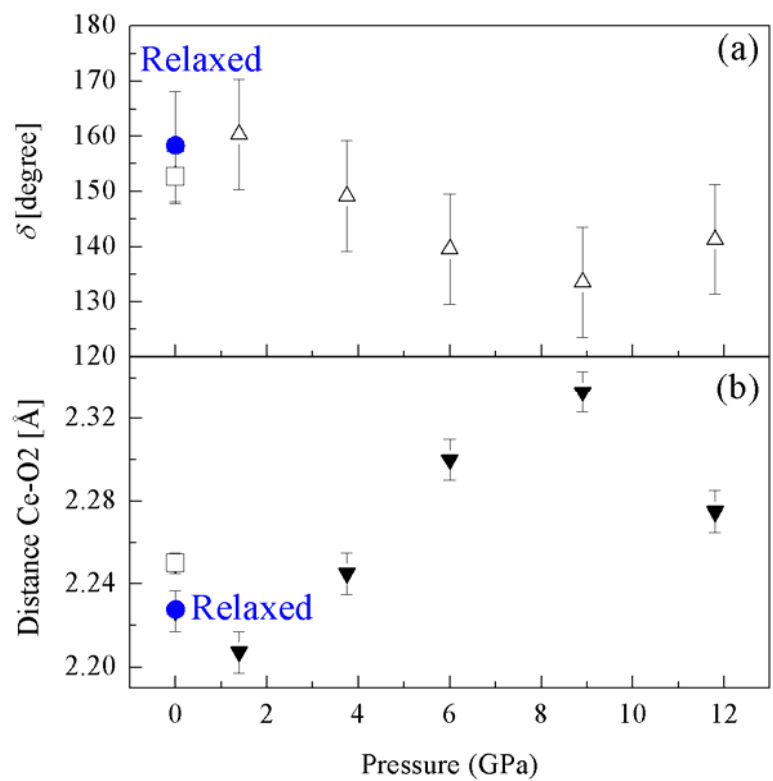

Figure 7. Evolution of the angle between Ce-O2-Ce ($\gamma$) for hydrated BaCe$_{0.8}$Y$_{0.2}$O$_{3-\delta}$ (a) and distance between Ce and O2 (b) as a function of pressure.